\documentclass[prb,twocolumn,groupaddress]{revtex4}

\usepackage{amsmath} 
\usepackage{amssymb} 
 \usepackage{amsfonts}

\usepackage{graphicx} 

\usepackage{array}
\usepackage{multirow}
\begin{document}

\title{Rationalizing polymer swelling and collapse under repulsive to highly attractive cosolvent conditions} 

\author{Anja Muzdalo, Jan Heyda, and Joachim Dzubiella}
\affiliation{Soft Matter and Functional Materials, Helmholtz-Zentrum Berlin, Hahn-Meitner Platz 1, 14109 Berlin, Germany}
\affiliation{Department of Physics, Humboldt-University Berlin, Newtonstr.~15, 12489 Berlin, Germany}

\thanks{To whom correspondence should be addressed. E-mail: joachim.dzubiella@helmholtz-berlin.de}

\begin{abstract}
  
 The collapse and swelling behavior of a generic homopolymer is studied using implicit-solvent, 
explicit-cosolvent Langevin dynamics computer simulations for varying interaction strengths. The 
systematic investigation reveals that polymer swelling is maximal if both 
monomer-monomer and monomer-cosolute interactions are weakly attractive. In the most swollen 
state the cosolute
density inside the coil is remarkably bulk-like and homogenous. 
Highly  attractive monomer-cosolute interactions, however, are found to induce a  collapse  of the chain which, in contrast to the collapsed case induced by purely repulsive cosolvents, exhibits 
a considerably enhanced cosolute density within the globule. Thus, collapsed states, although 
appearing similar on a first glance, may result from very different mechanisms with 
distinct final structural and thermodynamics properties. Two theoretical models, 
one based on an effective one-component description where the cosolutes have been 
integrated out, and a fully two-component Flory -- de Gennes like model, support the 
simulation findings above and serve for interpretation.  In particular, the picture is supported 
that collapse in highly attractive cosolvents is driven by crosslinking-like bridging effects, while the ratio of attraction width to cosolute size plays a critical role behind this mechanism. Only if polymer-cosolute interactions are not 
too short-ranged swelling effects should be observable. Our findings may be important
for the interpretation of the effects of cosolutes on polymer and protein conformational 
structure, in particular for highly attractive interaction  combinations, such as provided
by urea, GdmCl, NaI, or NaClO$_4$ near peptide-like moieties.
   
\end{abstract}

\maketitle

\section{Introduction}

Single polymer coils in solution swell in good solvents and collapse 
in bad solvents.~\cite{flory} In a good solvent the effective interaction
between polymer monomers is repulsive  which tends to swell the
polymer. In a bad solvent the interaction is essentially attractive and the 
coil shrinks until hindered by steric packing effects. On the simplest
theoretical level these trends  can be qualified by effectively one-component
mean-field  treatments as pioneered by 
Flory and de Gennes.~\cite{flory, degennes}
Here the system free energy $F(R)$ in dependence of coil size $R$ is
typically written as~\cite{degennes, rubinstein}
\begin{eqnarray}
F(R) \sim  {R^2}/{Nb^2} + {B_2 N^2}/{R^3} + {B_3 N^3}/{R^6}, 
\label{flory}
\end{eqnarray}
where the first term is the elastic energy of an ideal chain with $N$ monomers
and segment length $b$, and the
other terms account for mutual monomer interactions globally 
expressed by an virial expansion up to second order in 
density $\sim N/R^3$. The virial coefficients have to be considered
as based on {\it effective} pair potentials because the (co)solvent degrees of freedom
have been integrated out. Scaling law predictions from (1) are basis for the discussion 
of collapse and swelling in polymer science.~\cite{degennes, rubinstein}

The problem of such a perspective is that the information about 
the detailed effects of solvent or cosolvents are lost. In the recent
years, however, the complexity and resolution of experimental 
investigations has increased, and there is growing interest about how the specific 
binding mechanisms of cosolutes may alter global polymeric 
properties, such as the coil-globule transition.~\cite{baysal}
A particularly important example is the specific effect 
of cosolutes, such as ions, osmolytes, or denaturants on protein structure 
and stability.~\cite{scholtz, dagget, bolen, obrien, cremer:review,record:pnas,record:urea, angel1, angel2, pavel1,pavel2} Here, general 
polymer  principles are typically employed as  a starting point for the investigation of those cosolute 
effects.~\cite{dill1,dill2}  In this spirit, relatively simple thermosensitive polymers such as the poly(N-isopropylacrylamide) (PNIPAM)  homopolymer often serve as a experimental peptide model.~\cite{pnipam:lcst,pnipam:urea} On the computational side,  simulations  allow to address fundamental questions 
in the framework of polymer physics, such as swelling and collapse of 
idealized, purely hydrophobic homopolymer  in denaturants, osmolytes,  
or salts,~\cite{berne:urea,garde,garde2,garde3}  or more generic 
polymer-cosolute systems.~\cite{polson,lowe,antypov}

An important experimental reference to our study are the works by the Cremer group on 
(methylated) urea and ion-specific  effects on the lower critical solution temperature 
(LCST) of PNIPAM~\cite{pnipam:urea,pnipam:lcst} and the relatively simple, 
elastin-like peptides.~\cite{pnipam:urea,elastin} The qualitative change in the LCST, $\Delta T(\rho)$, 
with cosolute concentration $\rho$ can be used as an index whether polymer coils swell or 
collapse with the addition of cosolutes.~\cite{cremer:review} For instance, PNIPAM 
swells in methylated urea, while it collapses with the addition of nonmethylated urea. In contrast, 
the relatively hydrophobic elastin-like peptides swell in urea, 
consistent with computer simulations of simple homopolymers.~\cite{berne:urea,garde,garde2,garde3}
For both PNIPAM and elastin peptides, both being electroneutral, collapse and 
swelling in salt strongly depends on salt type. Hence, there is a wide variety of observed 
behaviors which are challenging to categorize.

Typically polymer collapse (or protein stabilization) is argued to originate from the preferential 
exclusion of repulsively interacting cosolutes, thereby trying to minimize cosolute-accessible
surface area.~\cite{bolen,record,record:pnas,record:urea}  Sagle {\it et al.}, however, argue 
that urea collapses PNIPAM due to a direct binding mechanism featuring strong, short-ranged 
attractions provided by multivalent hydrogen bonds.~\cite{pnipam:urea} Crosslinking then 
leads to shrinking of the coil. This were not the case,  they say, for methylated urea, or urea 
interacting with peptides, where interactions are only monovalent and very weakly attractive.
Hence, specific values of binding parameters such as attraction strength and width  are probably 
decisive whether a given polymer swells or not. 

The crosslinking effect may also be important for other cosolutes or salts, such as 
NaClO$_4$.  Its effects on the LCST of PNIPAM does not conform with the usual 
Hofmeister series.~\cite{pnipam:lcst, cremer:review} On the other hand, it is known to strongly 
interact with the peptide group.~\cite{pnipam:urea,pnipam:lcst} Indeed recent 
explicit-water computer simulations together with circular dichroism (CD) 
and F{\"o}rster resonance energy transfer (FRET) measurements on 
NaClO$_4$-destabilized $\alpha$-helical peptides demonstrated highly compact, 
disordered states crosslinked by a network of sodium and perchlorate ions.~\cite{alvaro} Thus, collapsed states,  although appearing similar on  a first glance, may result from very different mechanisms with quite distinct final structural and 
thermodynamics properties. Indications of the latter stem from the experimental characterization 
of NaClO$_4$ denatured  molten globules~\cite{hamada, scholtz, alvaro}, which may be cosolute rich, 
with implications in protein folding.~\cite{haran,dryglobule} We note that similar considerations 
may be important in the ion-induced  collapse of polyelectrolytes by ion condensation 
beyond simple Debye-H{\"u}ckel electrostatics.~\cite{winkler,khoklov,schiessel, mutu}
Hence, finding minimalistic models which can describe this vast variety
of effects are in need.

The aim of this paper is to theoretically investigate polymer collapse and swelling 
under the action  of cosolutes on a highly generic level. For that we first start to 
employ implicit-solvent computer simulations of a Gaussian polymer chain including explicit
cosolutes, and we systematically vary monomer-monomer and monomer-cosolute 
attraction strengths. Our polymer model is similar to that of Toan {\it et al.}~\cite{thirumalai} 
who systematically  investigated polymer swelling and collapse (with consequences to 
conformational kinetics) in  dependence of monomer-monomer attraction strength in purely implicit solvent.
Our simulations indicate chain collapse for highly attractive cosolute conditions as also found in recent on- and off-lattice 
computer simulations by Antypov and Elliot.~\cite{antypov} The results from the simulations are then 
rationalized by two theoretical models: First, an effective interaction model where the
action of the monomer-cosolute is explicitly integrated out in  a statistical mechanics 
framework.~\cite{weikl} The results are then discussed on a one-component level as in eq~(1). 
Secondly, we extend the Flory-deGennes description in eq.~(1) to a full 2-component 
description and calculate swelling behavior in dependence of interaction strengths and 
cosolute density. Both theoretical descriptions qualitatively agree with the simulations
giving important rational. Importantly, all three approaches point to polymer collapse
at highly attractive cosolute conditions due to crosslinking-like bridging effects as
argued by Sagle {\it et al.}~\cite{pnipam:urea} 
 
\section{Polymer-Cosolute Langevin Computer Simulation}

\subsection{Model and Methods}

In our simulation we consider a single, coarse-grained homopolymer with $N=100$ connected monomers 
in an implicit solvent background in a volume V. In addition, $N_c$ explicit cosolutes
with mean number density $\rho_c^0 = N_c/V$ are added to the system.
All particles, polymer monomers and cosolutes, are interacting via the Lennard-Jones (LJ) 
pair potential 
\begin{eqnarray}
V_{ij}(r) = 4\epsilon_{ij}[ ({\sigma_{ij}}/{r})^{12} - ({\sigma_{ij}}/{r})^6],
\label{LJ}
\end{eqnarray}
where $i = m,c$ stands for monomer or cosolute atom, respectively. The value of the size 
$\sigma_{ij}\equiv\sigma=0.3385$~nm is chosen to be be fixed for all interactions in our simulations. 
It is the same value as in Toan {\it et al.}'s work~\cite{thirumalai} and is comparable to the typical  size of the 
methyl group.  The interaction energy $\epsilon_{ij}$ will be 
systematically varied as described below. The polymer is modeled by 
a Gaussian chain with harmonic nearest-neighbor bond interactions $V_{{\rm bond},ij}=0.5k({\bf r}_i-{\bf r}_j)^2$
and a spring constant $k=320$~kJ mol$^{-1}$ nm$^{-2}$. The latter is chosen such that we end up with
the same effective Kuhn length $b=0.38$~nm for the ideal polymer as in Toan {\it et al.}'s work.~\cite{thirumalai} 
The mean end-to-end distance in the ideal case is thus $\bar R = \sqrt{N}b=3.8$~nm. The corresponding
ideal radius of gyration $R_g^0=1.55$~nm.  The  LJ interaction between a monomer and its next two nearest 
neighbors is excluded.

The polymer-cosolute system 
is simulated in the $NVT$-ensemble 
using stochastic computer simulations.  Every atom is propagated via the Langevin equation
\begin{eqnarray}
m\ddot{\bf r}_i + m\xi \dot{\bf r}_i = \sum_{j}{\bf F}_{ij} + {\bf F}^{(R)},
\label{BD}
\end{eqnarray} 
where ${\bf r}_i$ is the position of a particle $i$, $m=8$~amu its (irrelevant) mass, $\xi=0.5$~ps$^{-1}$ 
is the friction constant, ${\bf F}_{ij}$ the force between particles $i$ and $j$, 
and ${\bf F}^{(R)}$ the stochastic force stemming from the solvent kicks. The stochastic
force has zero mean $\langle{\bf F}^{(R)}\rangle=0$ and fluctuation-dissipation is obeyed 
via $\langle{\bf F}_i^{(R)}(t){\bf F}_j^{(R)}(t')\rangle=2 k_BT \xi\delta_{ij}\delta(t-t')$. 
The simulations are performed using the Gromacs simulation package~\cite{gromacs} with periodic
boundary conditions,  an integration time step of 4 fs, and lincs bond constraints.
The simulation box is  cubic and has a box length of 13 nm. We fix the mean cosolute density 
$\rho_c^0=$~3~nm$^{-3}$ ($\simeq 5$~mol/l) chosen to represent a typical denaturant density. 
With that we end up with $N_c=6591$ cosolutes in the simulation. We simulate every system 
for at least $1~\mu$s.

If the polymer is in  a swollen or collapsed state can be judged by inspection of the
radius of gyration $R_g$  scaled by the one of the ideal polymer $R_g^0$. 
If $R_g/R_g^0>1$ the polymer is swollen, if $R_g/R_g^0<1$, it is collapsed. The statistical error of $R_g$ 
in our simulations is estimated by block averages: the trajectory is divided into $n$ blocks of the same length 
(at least 200ns, but up to 1$\mu$s) and the standard deviation of these block averages 
with respect to the mean was calculated and divided by $\sqrt{n}$ to obtain the error.

Whether the LJ interaction $V_{ij}(r)$ is globally attractive or repulsive can be quantified by 
the second virial coefficient  defined via 
\begin{eqnarray}
B_2^{ij} = -\frac{1}{2} \int {\rm d}^3r \left[\exp(-\beta V_{ij}(r))-1\right],
\label{B2}
\end{eqnarray} 
where $\beta = (k_BT)^{-1}$ is the inverse thermal energy and we use $k_BT=1$ as the energy scale
in the following.  For the LJ potential, the $B_2$ in dependence of the LJ interaction energy $\epsilon$ 
is shown in Fig.~1: for interaction
values $\epsilon \lesssim 0.3$, the $B_2$ is larger zero since excluded volume contributions 
from distances $r<\sigma$ dominate. At $\epsilon = 0.3$ the $B_2$ vanishes, i.e., interactions are ideal on 
a 2-body level. For $\epsilon \gtrsim 0.3$ effectively the interactions are attractive. In the polymer-only
case (no cosolutes), those regimes can be identified with the usual good solvent regime (swelling), 
$\Theta$-solvent (ideal), and bad solvent (collapse), respectively.~\cite{rubinstein} Indeed previous simulations 
showed a collapse transition of the 100mer in the
weakly attractive regime with $\epsilon$ around 0.4. For $\epsilon\gtrsim0.6$
the chain collapses in dense, compact states. This regime was characterized as 
highly attractive.~\cite{thirumalai}

In our simulations we vary systematically both $\epsilon_{mm}$ and $\epsilon_{mc}$ between 0 and 1, thereby 
studying the whole range of repulsion to strong attraction for both components, that is, monomer-monomer 
and monomer-cosolute interactions.  For simplicity the cosolute-cosolute interaction will be 
fixed to $\epsilon_{cc}=0.3$ mimicking a near-ideal cosolute solution. Nonideality may also 
influence  polymer swelling and shrinking behavior due to osmotic effects but these effects
are out of scope of this paper. 

\vspace{1cm}
\begin{figure}[h]
\begin{center}
\includegraphics[width=8.0cm,angle=0]{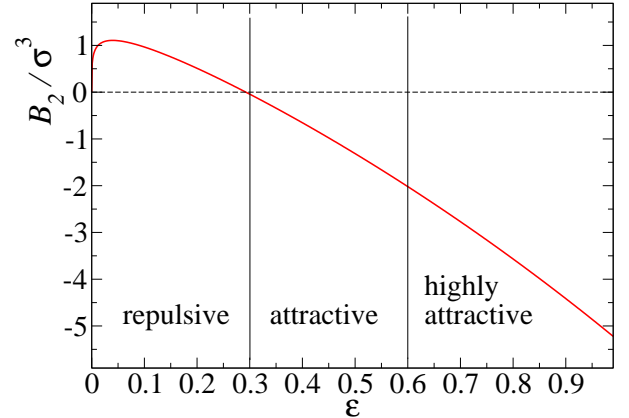}
\caption{Second virial coefficient eq.~(\ref{B2}) of the Lennard-Jones interaction eq.~(\ref{LJ}) versus 
LJ interaction energy $\epsilon$. The $B_2$ value can be used to classify the interaction
into mainly 'repulsive' ($B_2>0$), 'attractive' ($B_2<0$), and 'highly attractive' 
($B_2/\sigma^3\lesssim2$).}
\end{center}
\end{figure}

\subsection{Simulation Results}

In Fig.~2 we plot the size of the polymer expressed by the radius of squared gyration $R_g^2$  versus monomer-cosolute attraction $\epsilon_{mc}$ for varying monomer-cosolute 
attraction $\epsilon_{mm}$.  Note that $R_g^2$ is scaled by $(R_g^0)^2$, the size of the ideal
polymer, to easily distinguish between swollen ($R_g/R_g^0>1$) and collapsed 
($R_g/R_g^0<1$) states.  Let us first focus on vanishing $\epsilon_{mc}=0$, i.e.,  the polymer-only 
case: the polymer is swollen in a good solvent $\epsilon_{mm}\lesssim0.3$, almost ideal 
for $\epsilon_{mm} = 0.4$, and collapsed for $\epsilon_{mm}\gtrsim 0.5$.  

\vspace{1cm}
\begin{figure}[h]
\begin{center}
\includegraphics[width=7.0cm,angle=0]{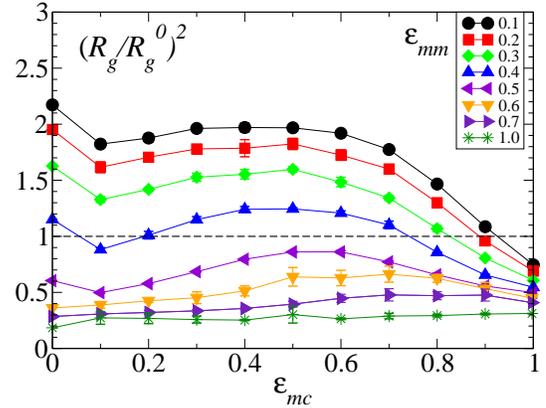}
\caption{The squared radius of gyration $R_g^2$ of the polymer scaled by the ideal value $(R_g^0)^2$ versus the monomer-cosolute
interaction strength $\epsilon_{mc}$ for varying monomer-monomer interaction $\epsilon_{mm}$. If $(R_g/R_g^0)^2>1$ the 
polymer is swollen, if $(R_g/R_g^0)^2<1$, it is collapsed. Error bars are typically of
symbol size.}
\end{center}
\end{figure}

If now  the cosolutes are 'switched-on'  
($\epsilon_{mc}>0$)  to a value of $\epsilon_{mc}=0.1$, where monomer-cosolute
interactions are repulsive (cf. Fig.~1),  the chain shrinks considerably for $\epsilon_{mm}<0.6$. 
This is expected  since the exclusion of repulsive monomers from the polymer region is entropically 
disfavored and leads to chain shrinking. For larger $\epsilon_{mm}>0.6$ the effects are small. 
For an increased $\epsilon_{mc}=0.2$ the polymer swells 
for all of the values of $\epsilon_{mm}$.  Hence, the solvent quality overall improves with less 
monomer-cosolute repulsion (cf. Fig.~1). For very strong monomer-monomer attractions, 
$\epsilon_{mm}>0.6$, the effect is quite small and the polymer stays essentially 
in the collapsed state. 

For further increasing $\epsilon_{mc}\gtrsim 0.3$, roughly where monomer-cosolute 
repulsion turns into attraction, the swelling of the polymer continues for all $\epsilon_{mm}$ to a maximum 
value $R_g^{max}$ whose   corresponding $\epsilon_{mc}^{max}$ value depends on $\epsilon_{mm}$.  For larger monomer-monomer 
attraction $\epsilon_{mc}^{max}$ is shifted to larger values.  For 
values larger than $\epsilon_{mc}^{max}$ the chain collapses again, i.e., there is a reentrant
collapse transition for an increasingly attractive monomer-cosolute interaction. This continuous crossover 
from a good solvent to a bad one with increasing attraction  has been observed 
already in a previous simulations.~\cite{antypov}  For not highly attractive
 intrapolymer interactions $\epsilon_{mm}<0.6$ and highly attractive polymer-cosolute
interactions $\epsilon_{mc}>0.6$, the chain can be even more strongly collapsed than 
in the case of highly repulsive monomer-cosolute ($\epsilon_{mc}=0.1$) interactions. 
Hence, this regime shows strong compaction by a highly attractive cosolute.

The cosolute-induced effect is strongest for $\epsilon_{mm}\simeq0.4$, that is, 
in the ideal to weakly attractive polymer regime. Overall the polymer crosses from 
collapsed ($\epsilon_{mc} \simeq 0.1$) to swollen ($\epsilon_{mc} \simeq 0.3-0.8$)
to again collapsed ($\epsilon_{mc} \gtrsim 0.8$) states with the most dominant changes when compared to the other monomer-monomer interactions. Simulation snapshots of the polymer-cosolute system
for $\epsilon_{mm}\simeq0.4$ and $\epsilon_{mc}=$0.1 (collapsed), 0.5 (swollen), and 
1.0 (collapsed)  are shown in Fig.~3. Note the considerably different amount of cosolutes 
in the vicinity  of the two collapsed states (a) and (c). 

\begin{figure}[h]
\begin{center}
\includegraphics[width=8.5cm,angle=0]{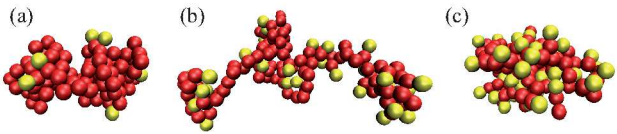}
\caption{Simulation snapshots of the polymer-cosolute system
for $\epsilon_{mm}\simeq0.4$ and (a) $\epsilon_{mc}=$0.1 (collapsed), (b) 0.5 (swollen), and (c) 
1.0 (collapsed). Cosolutes (yellow spheres) are shown which are found in a cut-off distance of 
$\sigma2^{1/6}=3.8$\AA~ to the polymer backbone (connected red spheres).}
\end{center}
\end{figure}

To further characterize structural details in the regimes described above we plot the radial 
one-particle density distributions of the monomers and cosolutes 
$\rho_m(r)$ and $\rho_c(r)$ in Fig.~4. With $r$ we denote the radial distance to the
center-of-mass of the polymer chain. The profiles are plotted for a 
fixed $\epsilon_{mm}=0.4$ and varying $\epsilon_{mc}=0.1$, 0.5, and 1.0.  
In the collapsed states the polymer density profiles are similar. However, the
cosolutes are strongly depleted in the repulsive case, $\epsilon_{mc}=$0.1, while
their presence is massively enhanced in the highly attractive case with
$\epsilon_{mc}=$1.0. At the most swollen polymer state for  $\epsilon_{mc}=0.5$ 
the cosolute density is remarkably homogeneous and bulk-like. 

\begin{figure}[h]
\begin{center}
\includegraphics[width=8.0cm,angle=0]{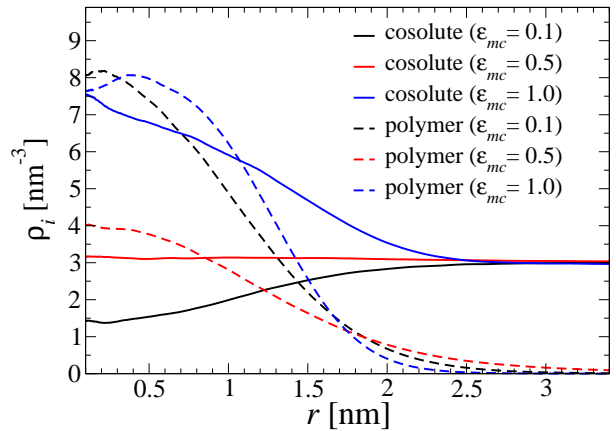}
\caption{Radial density profiles $\rho_i(r)$ of polymer monomers ($i=m$) and cosolutes $(i=c)$ around
the center-of-mass of the polymer. Interaction values are $\epsilon_{mm}=0.4$ and $\epsilon_{mc}=0.1$, 0.5, and 1.0
(see legend) corresponding to the snapshots in Fig.~3.}
\end{center}
\end{figure}

\section{Effective Interaction Model}

In this section we attempt to rationalize some of the swelling and collapse trends we found above 
in the perspective of an effective one-component model. For this the effects of the
cosolutes have to be integrated out to derive an effective cosolute-induced 
interaction between two monomers. This is reflected then in an effective $B_2$ coefficient as
used in the one-component Flory approach eq (\ref{flory}).  For simplicity we will model the monomers
as planar plates and aim only at qualitative statements. We borrow thereby from a similar model introduced 
within the framework of  mesosurface attraction induced by adhesive particles.~\cite{weikl}

\begin{figure}[h]
\includegraphics[width=7cm,angle=0]{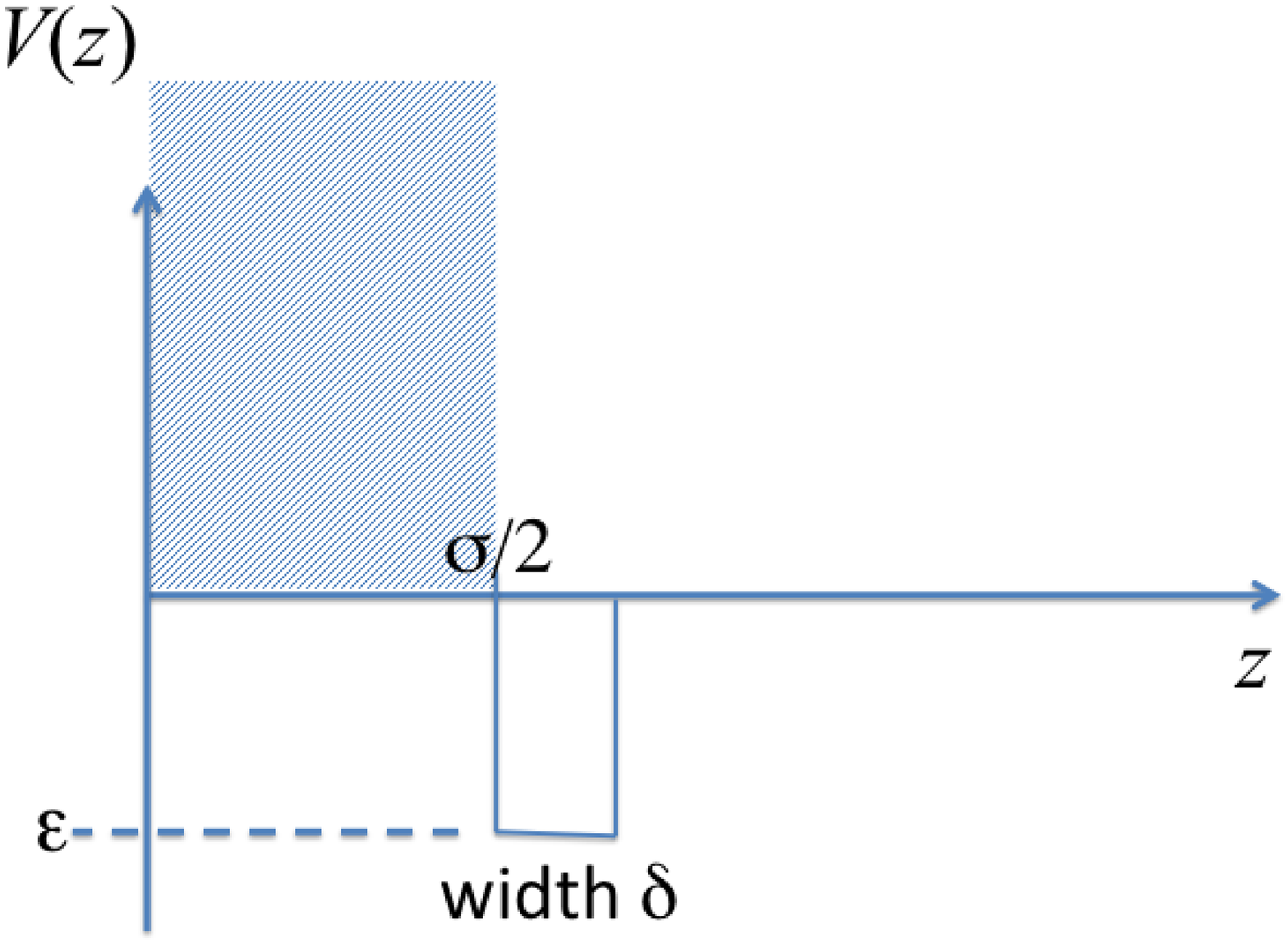}
\includegraphics[width=6.5cm,angle=0]{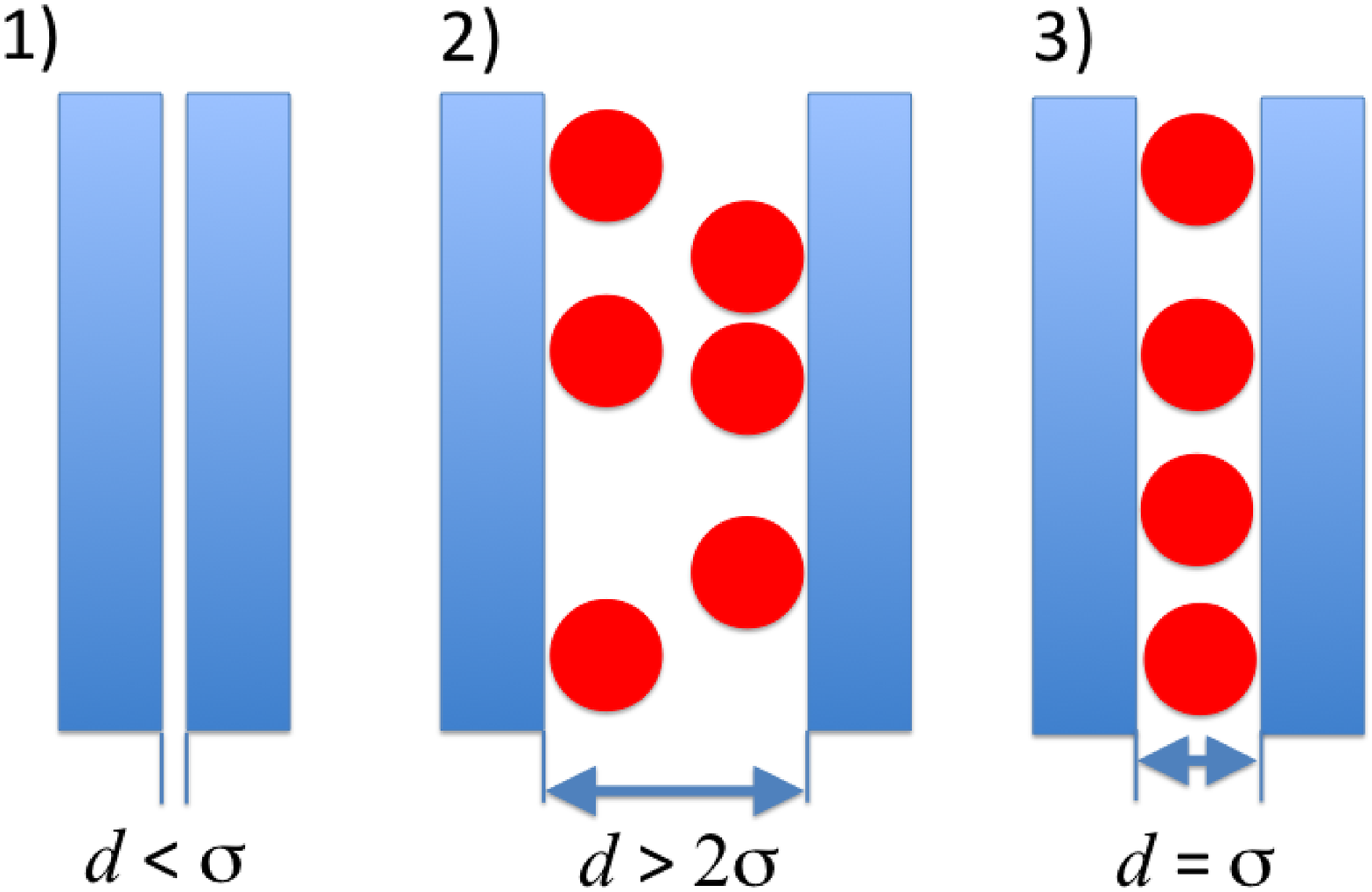}
\caption{Top: Generic interaction potential $V(z)$ between one cosolute and one plate-like monomer in a center-of-mass to
 surface distance $z$ as assumed in our statistical mechanics approach. $\sigma/2$ is the cosolute hard-core radius, 
 while  $\epsilon<0$ is the attractive well depth and $\delta$ is its width. 
Bottom: sketch of the three interaction situations of the simplified plate-like monomers (blue plates) 
in presence of the cosolutes (red spheres).}
\end{figure}

Consider two planar surfaces (monomers) with area $A$ in a surface-to-surface distance $d$ in contact with a reservoir of 
cosolutes at concentration $\rho_c^0$.  The  cosolutes interact with the surface 
with the generic potential $V(z)$ as shown in Fig.~5: the cosolutes are hard spheres with a diameter $\sigma$ 
and have an additional attractive (adsorption) energy of strength $\epsilon\leq0$  and not too large width $\delta \lesssim \sigma/2$. 
Now consider three situations 1)-3) as also depicted in  Fig.~5. In situation 1)  the plate distance is  $d<\sigma$, i.e, 
no cosolutes fit between them. This situation can be coined 'depletion' situation. In situation 2), $d>2\sigma$, and cosolutes can solvate 
both of the surfaces.  In situation 3), $d\simeq \sigma$, one bound particle can interact with the two 
surfaces simultaneously. This situation can be named 'bridging' or 'crosslinked' 
situation. Interactions between the cosolutes, e.g., packing effects,  
are neglected.  These effects have been included in a more complete calculation within a 
similar model previously.~\cite{weikl}

In case 1) the proximity of the plates increases the accessible volume for the cosolutes. 
The latter gain configurational entropy proportional to the freed volume $A\sigma$. This leads
to a favorable (grand canonical) free energy per area 
\begin{eqnarray}
\Omega_1^*/A = -\sigma \rho_c^0
\end{eqnarray}
 for $d<\sigma$.  That is the well-known effective surface attraction by particle depletion.~\cite{hunter} 
The grand canonical free energies for situations 2) and 3) can be explicitly written down and are 
\begin{eqnarray}
\Omega_2 = N_b\left[\ln\left(\frac{N_b\Lambda^3}{2A\delta}\right)-1\right] + N_b(\epsilon-\mu)
\end{eqnarray}
and
\begin{eqnarray}
\Omega_3 = N_b\left[\ln\left(\frac{N_b\Lambda^3}{A\delta}\right)-1\right] + N_b(2\epsilon-\mu),
\end{eqnarray}
respectively, where  $\Lambda$ is the thermal wavelength, $\mu$ the chemical potential,  and $N_b$ is the number of bound particles. 
Note the difference between (2) and (3): In case 2), a particle gains energy $\epsilon$ by binding to one surface, but 2 surfaces, 
that is, twice the configuration space as in 3) is available in total (hence the factor 2 before $A$ in the log-term). In situation 3), one particle 
gains $2\epsilon$ by binding to two surfaces simultaneously, but effectively only one plane for the cosolute translation is available. 
The energies are minimized by the 
Boltzmann equations ${N}/(2\delta A) = \rho_c^0\exp(-\epsilon)$ and
${N}/(\delta A) = \rho_c^0\exp(-2\epsilon)$, respectively, 
where we used $\rho_c^0=\Lambda^{-3}\exp(\mu)$. Plugging the solution back in (1) and (2) 
we obtain the minimum free energies per area
\begin{eqnarray}
\Omega_2^*/A = -2\delta\rho_c^0\exp(-\epsilon)
\label{omega2}
\end{eqnarray}
 and
\begin{eqnarray}
\Omega_3^*/A = -\delta\rho_c^0\exp(-2\epsilon).
\label{omega3}
\end{eqnarray}

Inspection of the results shows that for very small $|\epsilon|$ situation 1) is favored over 2), that is, 
the depletion interaction wins. For increasing $\epsilon$, however, situation
2) is favored over 1); when comparing $\Omega_1^*=\Omega_2^*$ we find  
$\epsilon_{12}^* = - \ln {\sigma}/(2\delta)$ which depends on the interaction range $\delta$. 
For a LJ interaction, as in our simulation model, the width of the minimum roughly 
$\delta\simeq \sigma/3$ which yields  $\epsilon_{12}^* \simeq - 0.4$ for the crossover from 
depletion attraction to full solvation of the two monomers. Since the minimum free energies of the two situations
1) and 3), where $d\lesssim\sigma$, are both higher, this implies repulsion between close monomers. In colloidal
physics this effect was coined 'repulsion through attraction'.~\cite{ard,stelios}  The free energy for situation 3), 
where  bridging is favored, decreases faster  with $\epsilon$ than in situation 2), because of the 2 in the exponent. 
By equating (\ref{omega2}) and (\ref{omega3}), we  obtain  the crossover energy 
$\epsilon_{23}^* = -  \ln 2\simeq -0.7$.  For high attraction consequently situation 3) 
is favored.

Thus, we can now generally identify three regimes depending on the attractive interaction strength $\epsilon$: i) $|\epsilon|\lesssim 0.4$: depletion attraction between surfaces (favoring collapse); ii)  $0.4\lesssim |\epsilon| \lesssim 0.7$: repulsion 
through attraction (favoring swelling); $|\epsilon| \gtrsim 0.7$: polymer-polymer attraction by bridging (favoring collapse). The situation is summarized in the 'phase diagram' in Fig.~6 for the specific  value of $\delta=\sigma/3$. Note, however,  that the attraction width 
plays a nontrivial role:  if the attraction range 
is small compared  to the cosolute size, $\delta < \sigma/4$, then $|\epsilon_{12}^*| > 0.7$, and situation 2)  
becomes metastable.  In contrast, if $\delta$ is not small, $\delta \geq \sigma/2$, situation 2 is favored over 1) 
for any value $\epsilon>0$ and the depletion scenario 1) becomes metastable. However, for a large 
attraction width, $\delta>\sigma/2$, the bridging case is  ill-defined and is possibly indistinguishable
anymore from the solvated case 2).  We note that   
equivalent scenarios apply in mixtures of colloids and nanoparticles.~\cite{stelios, scheer}

\vspace{1cm}
\begin{figure}[h]
\includegraphics[width=7.0cm,angle=0]{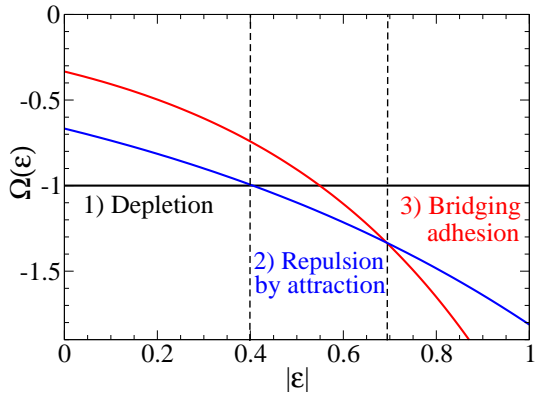}
\caption{ Free energy branches for the situations 1)-3) depicted in Fig.~5 versus the cosolute-monomer 
interaction energy $\epsilon$. Parameters chosen here are density $\sigma^3\rho_c^0=1$ and 
interaction width $\delta=\sigma/3$.}
\end{figure}

The  findings above may be important to understand controversial effects observed in experimental LCST
measurements of polymers and simple peptides. Only if polymer-cosolute attractions are weak and 
not too short-ranged, swelling effects 
should be observable. This may be indeed valid for weak hydrophobic or dispersion attractions.
In computer simulations in fact urea swells purely hydrophobic polymers.~\cite{nico:urea,berne:urea} 
In  hydrogen bonding  systems, however,  the attraction length is short $\simeq 0.1-0.2$~nm and no swelling is 
possible. Consistent with that view, urea collapsed PNIPAM polymers which has been argued is 
due bridging by short-ranged H-bonds, strongly supported by experimental means.~\cite{pnipam:urea}
The same experiments also showed urea-induced swelling of hydrophobic peptides. Consistent
with our picture it was argued that this is due to much weaker, not short-ranged 
attractions.~\cite{pnipam:urea}

Looking at the quantitative numbers predicted by the present theory they are surprisingly close
to the ones observed in our simulations. The chain was found to be most swollen in 
the weakly attractive regime $\epsilon\simeq0.3-0.6$ for not too high values of 
intrapolymer attraction. Thus although highly simplified, the effective one-component
approach seems to capture most of the underlying physics.

\section{Two-component Flory - de Gennes model}

Another perspective to polymer collapse in a highly attractive cosolute dispersion
could be based on a mean-field Flory-de Gennes picture where all interactions
are described by 2nd and 3rd order virial coefficients. In contrast to an effective
one-component model as typically used in literature, cf. eq.~(\ref{flory}), 
we now investigate the full two-component description. 

Consider a polymer with $N$ monomers and an effective bond length $b$ in contact with 
a reservoir of cosolutes with density $\rho_c^0$. As usual in Flory theory we take now the end-to-end
distance $R$ to represent the size of the polymer coil. In this view the monomer density profile 
is just a step function with a constant monomer number density $\rho_m=N/V=3N/(4\pi R^3)$ inside
the coil with volume $V$.  In the spirit  of mean-field Flory theory for real chains the excess free energy due to 
interactions can be expressed in terms of the virial expansion in the monomer density $\rho_m$ 
and the cosolute density $\rho_c$ within the coil.  A Flory-like free energy of the system can be 
then written as
\begin{eqnarray}
\label{flory2}
F(R) &=& \frac{3R^2}{2Nb^2} + \frac{\pi^2Nb^2}{12R^2} \\&+& V\sum_{ij=m,c} \rho_i \rho_j B_2^{ij}+\frac{V}{2}\sum_{ijk=m,c}\rho_i\rho_j\rho_kB_3^{ijk}, \nonumber
\end{eqnarray}
where the first term denotes the elastic free energy of the ideal chain, the second term is a correction due 
to confinement entropy for highly collapsed states,~\cite{doi:edwards,rubinstein} and the last two terms are the virial 
corrections up to third order. The density $\rho_c=\rho_c(R)$ denotes  the cosolute density inside the polymer 
which is related to the bulk concentration $\rho_c^0$ by  
\begin{eqnarray}
\rho_c(R) = \rho_c^0\exp(-\mu_{\rm exc}) 
\end{eqnarray}
with the the excess chemical potential
\begin{eqnarray}
\mu_{\rm exc} = \frac{\partial}{\partial \rho_c}\left[\sum_{ij=m,c} \rho_i \rho_j B_2^{ij}+\frac{1}{2}\sum_{ijk=m,c}\rho_i\rho_j\rho_kB_3^{ijk}\right]_V.
\label{muex}
\end{eqnarray}
The coupled eqs.~(\ref{flory2}) to (\ref{muex}) provide 
a  free energy expression as a function of $R$, the polymer size for given interactions
$V_{ij}$, with $i=c,m$, as provided by eq.~(\ref{LJ}),  polymerization $N$, 
bond length $b$, and  cosolute  concentration $\rho_c^0$. To obtain the equilibrium radius $R$ 
and corresponding monomer and cosolute densities we minimize (\ref{flory2}) according to 
$\partial F/\partial R=0$  numerically using an  iterative Newton-Raphson scheme. Due to the 
relative simplicity of the equations this takes only seconds on a single, ordinary computing 
processor.  

The $B_2$ values are explicitly calculated using definition (\ref{B2}). To connect properly to our 
computer simulation, where $\epsilon_{cc}=0.3$ and thus $B_2^{cc}$ vanishes, we set $B_2^{cc}=0$ 
accordingly. We have also explicitly calculated all values $B_3^{ijk}$ but found that  they
can be well approximated by a constant $B_3^{ijk} = 2\sigma^6$. The latter actually closely 
corresponds to the 3rd virial coefficient of hard spheres with diameter $\sigma$. In other 
words the  $B_3$ values reflect mostly packing effects by hard spheres with size $\sigma$.

\begin{figure}[h]
\begin{center}
\vspace{1.0cm}
\includegraphics[width=7.0cm,angle=0]{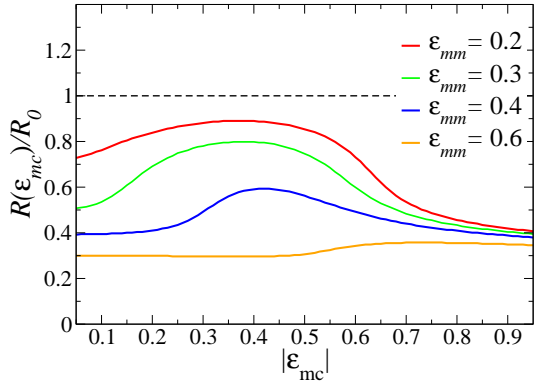}
\caption{The polymer radius $R$ scaled by its ideal value $R_0$ from minimization of the two-component Flory-de Gennes 
free energy eq.~(\ref{flory2}) versus polymer-cosolute interaction strength $\epsilon_{mc}$. The plot is for a polymer-polymer interaction $\epsilon=0.2$, 0.3, 0.4, and 0.6 . The polymer shows transitions from collapsed states 
to  swollen and collapsed states again , in qualitative
agreement with the Langevin computer simulations (cf. Fig.~2).}
\end{center}
\end{figure}

Results for the polymer size $R(\epsilon_{mc})$ are plotted in Fig.~7  
as a function of  $\epsilon_{mc}$ for $\epsilon_{mm}=0.2$, 0.3, 0.4, and 0.6 
(cf. curves in Fig.~2).  The radius is scaled by $R_0=3.27$~nm which is the ideal 
chain size according to the Flory approach.   As in the simulation the chain is collapsed 
for small $\epsilon_{mc}$ due the addition of repulsive cosolutes. 
Increasing $\epsilon_{mc}$ leads to chain swelling up to a maximum coil size with a corresponding 
$\epsilon_{mc}^{max}$ that increases with $\epsilon_{mm}$. This trend reproduces the one
found in the simulation. Further increase of $\epsilon_{mc}$ 
leads to  deswelling and reentrant collapse.  For the three smaller values of $\epsilon_{mm}$ the 
collapse transition is at about $\epsilon\simeq 0.7$.  This value is interestingly 
in agreement with the value obtained in the effective one-component approach
for the crossover from swelling to collapse. This may be accidental. However, 
all three approaches show the same qualitative features and numbers in the
same ballpark. Importantly, also the two-component Flory describes the
reentrant collapse in strongly attractive cosolute conditions. Note that the mean-field theory does not provide 'swelling' as defined by $R/R_0>1$.Trends, however, are, as discussed above, nicely in accord.

\begin{figure}[h]
\begin{center}
\vspace{1.0cm}
\includegraphics[width=7.0cm,angle=0]{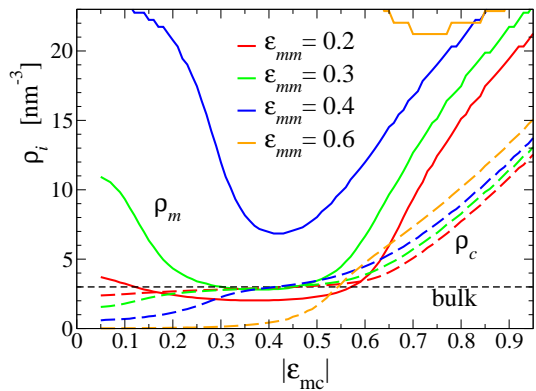}
\caption{The mean monomer density $\rho_m$ (solid lines) and cosolute density $\rho_c$ (dashed line) inside the polymer
coil versus monomer-cosolute attraction $\epsilon_{mc}$ and varying $\epsilon_{mm}=0.2$, 0.3, 0.4, and 0.6 (see legend). The curves are 
from the solution of the two-component Flory-like 
model for polymer collapse and swelling. For the three smaller $\epsilon_{mm}$ values,  
at the minimum of $\rho_m$ the cosolute density $\rho_c$ is comparable to 
its bulk value (horizontal dashed black line).}
\end{center}
\end{figure}

The mean monomer and cosolute densities $\rho_m(\epsilon_{mc})$ and $\rho_c(\epsilon_{mc})$, respectively, 
are plotted in Fig.~8 for  $\epsilon_{mm} = 0.2$, 0.3, 0.4, and 0.6 as a function of 
 $\epsilon_{mc}$.  The monomer 
density $\rho_m$ decreases and increases according to the swelling behavior 
$\rho_m\propto R^{-3}$ (cf. Fig. 7).  The mean cosolute density $\rho_c$ monotonically increases
with $\epsilon_{mc}$. For small values $\epsilon_{mc}<0.4$, cosolutes are
depleted from the polymer coil due to repulsive interactions. At $\epsilon_{mc}=0.4$, the density 
goes  down to one third  of the bulk density, comparable to the decrease found in the simulations, cf. Fig.~4. 
For the three smaller values of $\epsilon_{mm}$ at around $\epsilon_{mc}=0.4$, 
where the polymer is most swollen, there is a  plateau in the mean density which 
almost exactly corresponds to bulk density. Remarkably this reproduces the observation in 
the simulation profiles in Fig.~4, where we found a homogenous bulk-like density in 
the most swollen state of the coil. For the largest plotted $\epsilon_{mm}=0.6$ where
polymer compaction is very strong this correspondence between the minimum in 
$\rho_m$ and plateau $\rho_c$ is lost for whatever reason.
For further increasing  $\epsilon_{mc}\gtrsim 0.5$ the cosolute density $\rho_c$ within the 
polymer further grows for all values of $\epsilon_{mm}$ while the chain is collapsing again. 
Thus, consistent with the  simulations,  chain collapse driven by highly attractive polymer-cosolute 
interactions leads to highly dense cosolute states inside the coil. The density of about 14 nm$^{-3}$
is about twice as high as in the simulation, cf. Fig.~4, but the Flory prediction of 
one-order-of-magnitude increase inside the globule is qualitatively correct.

\begin{figure}[h]
\begin{center}
\vspace{1.0cm}
\includegraphics[width=7.0cm,angle=0]{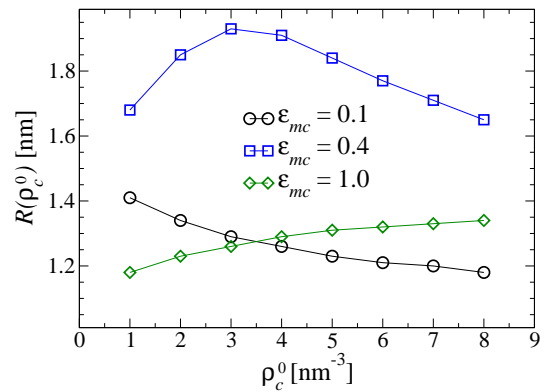}
\caption{Polymer size $R(\rho_c^0)$ in dependence of cosolute bulk density $\rho_c^0$ from the two-component 
Flory-like approach. Data are for a fixed monomer-monomer attraction $\epsilon_{mm}=0.4$ and varying 
monomer-cosolute attractions $\epsilon_{mc}=0.1$, 0.4, and 1.0. Note that only the case $\epsilon_{mc}=0.4$
exhibits nonmonotonic behavior.}
\end{center}
\end{figure}

Finally, we can use the Flory-like approach to make predictions about the dependence of 
polymer size $R$ on the cosolute density reservoir (or bulk) density $\rho_c^0$. 
The calculated $R(\rho_c^0)$ is plotted in Fig.~9 in dependence of cosolute bulk density $\rho_c^0$ 
for a fixed monomer-monomer attraction $\epsilon_{mm}=0.4$ and varying 
monomer-cosolute attractions $\epsilon_{mc}=0.1$, 0.4, and 1.0. The density range covers 
typical denaturant concentration from 1 to 8 nm$^{-3}$, that is, 1.7 to 13.3 mol/l. 
In the case $\epsilon_{mc}=0.1$, $R$ decreases monotonically with density. This 
is expected for a depletion-like collapse mechanism, where the effect grows with 
density. In contrast, in the case $\epsilon_{mc}=1.0$, $R$ {\it increases} monotonically with 
density. That indicates that the collapse effect due to highly attractive cosolute
deteriorates with increasing cosolute density. Remarkably, for $\epsilon_{mc}=0.4$, 
where swelling is found for weakly attractive cosolutes, the size behavior is 
{\it nonmonotonic.} This is interesting in the light of LCST measurements of 
poly(NIPAM) and elastin-like peptides in Hofmeister salts.~\cite{pnipam:lcst, elastin} 
Nonmonotonic  behavior of the LCST change with salt concentration 
was indeed observed only in the swelling  scenarios.  This happened for the salts 
NaI and NaSCN which seem to feature weakly attractive interactions with nonpolar  
peptide groups.~\cite{record}

\section{Summary and Concluding Remarks}

The coarse-grained computer simulations performed here and elsewhere~\cite{antypov}  
demonstrate polymer collapse to compact globular states at
highly attractive cosolute conditions. The polymer response and 
relative swelling/collapse behavior is most pronounced for polymers in nearly ideal 
$\Theta$-like solvent conditions. Here the polymer can shrink or 
swell considerably, depending on polymer-monomer
interaction strength. Collapsed states for repulsive and
strongly attractive cosolutes are physically different as 
in the latter state the polymer coil is rich in cosolutes. 
For the same polymer size the internal cosolute density 
can differ enormously between 1/2 to 1/3 of the bulk density (repulsive cosolutes) 
and  8 times the bulk density (highly attractive case), respectively, signifying considerably 
different physical properties of the globule-cosolute 
system.  The most swollen polymer state is characterized by a 
bulk like cosolute density inside the polymer coil. 

We further demonstrate that the transition to collapsed states
in strongly attractive cosolutes can be rationalized by two independent 
theoretical and semi-analytical  approaches.  Their implications are 
summarized in the following. 

The 'effective one-component
statistical mechanics model' showed that for large polymer cosolute attraction
bridging interactions (leading to 
compact  'glued' states) are energetically favored over both depletion 
interaction (favoring collapse by exclusion) and 
'repulsion by attraction' situation (leading to swelling). We have shown that 
a  critical parameter to distinguish those scenarios is the ratio between 
interaction width and cosolute size $\delta/\sigma$.  Only if polymer-cosolute 
interactions are not too short-ranged 
swelling effects  should be observable. 
This may be valid for hydrophobic or dispersion attractions.
In simulations indeed urea swells purely hydrophobic polymers~\cite{nico:urea,berne:urea} 
or hydrophobic peptides.~\cite{pnipam:urea}
In  hydrogen bonding  systems, however,  the attraction length is short 
$\delta/\sigma \simeq 1/4$ or 1/5 and no swelling is  possible. Consistent with that view, 
urea compresses PNIPAM polymers which has been argued is  due to H-bonding and 
crosslinking by bivalent binding as demonstrated by experiments.~\cite{pnipam:urea}
Methylated urea did not swell PNIPAM, neither did urea swell elastin-like peptides.

The two-component Flory-de Gennes mean-field model also predicts
polymer collapse at strongly attractive cosolute conditions. The transition 
from swollen to collapsed state at a polymer-cosolute attraction energy
of $-\ln 2~k_BT$ is in good agreement with our computer simulations and the
effective one-component approach.  In particular, the Flory model 
confirmed that cosolute densities inside the coil can significantly differ
between repulsive cosolutes and strongly attractive ones. It also reproduced
the simulation finding that the polymer is most swollen if internal cosolute
density exactly matches bulk density for nearly ideal cosolutes. 
Thus, the structural nature of collapsed states can be qualitatively
different while the polymeric state alone, on a first glance, may be
similar.  The Flory approach also predicts a nonmonotonic 
dependence of swelling magnitude with cosolute density, which
is observed for PNIPAM or elastin in NaI and NaSCN.~\cite{pnipam:lcst, elastin}
A more detailed investigation may shed more light on the reason why. 

We note that dense cosolutes in polymer coils 
may have strong impact on internal friction and viscosity of polymeric globules. 
Consequently, the 
conformational kinetics of polymers maybe highly affected
by explicit direct-binding cosolutes in contrast to isolated 
polymers.~\cite{thirumalai}

Our findings may have implications on the interpretation of denaturant 
action on proteins and the nature of denatured states. For instance, the
structure of  molten globular proteins  denatured by high NaClO$_4$ 
(sodium perchlorate) is controversially discussed.\cite{hamada, scholtz, alvaro}
Considering our results and the experimentally known strong affinity of
NaClO$_4$ to the protein backbone,~\cite{record,record:urea,record:pnas}
the reason for the difficulties may be a large amount of bridging or crosslinking 
denaturant which probably strongly obscures the experimental characterization 
of structure and its  classification.   

Finally, similar considerations as above may be important 
in the interpretation of ion-induced  collapse of polyelectrolytes beyond simple  
electrostatics.~\cite{winkler,khoklov,schiessel, mutu}

\acknowledgments

A.M. thanks the European Erasmus Programme for financial support. 
J.H. and J.D. acknowledge funding from the Alexander-von-Humboldt
Stiftung, Germany.

\end{document}